%% file: main.tex
\documentclass{article}
\usepackage{spconf}

\usepackage[pdftex]{graphicx}
\usepackage{amssymb}
\usepackage{amsmath}
\usepackage[noend]{algorithm2e}
\usepackage{times,balance,amsfonts,pgf,tikz,pgffor}
\usepackage{float} 
\usepackage[caption=false,font=footnotesize]{subfig}
\usepackage{color}
\usepackage{pgfplots}
\pgfplotsset{compat=1.7}

\begin{document}
\title{Hierarchical coded elastic computing}

\name{Shahrzad Kiani,
Tharindu Adikari, Stark C. Draper
\thanks{This research was supported in part by a Discovery Research Grant from the Natural Sciences and Engineering Research Council of Canada (NSERC), by DiDi scholarship, by the Ontario graduate scholarship, and by a grant from Huawei Technologies, Inc. This paper was presented at the IEEE International Conference on Acoustics, Speech and Signal Processing (ICASSP), Toronto, Canada, June 2021.}}
\address{Department of Electrical and  Computer Engineering, University of Toronto}

\maketitle

\begin{abstract}
Elasticity is offered by cloud service providers to exploit under-utilized computing resources. The low-cost elastic nodes can leave and join any time during the computation cycle. The possibility of elastic events occurring together with the problem of slow nodes, referred to as stragglers, increases the uncertainty of the system, leading to computation delay. Recent results have shown that coded computing can be used to reduce the negative effect of elasticity and stragglers. In this paper, we propose two hierarchical coded elastic computing schemes that can further speed up the system by exploiting stragglers and effectively allocating tasks among available nodes. In our simulations, our scheme realizes $45\%$ improvement in average finishing time compared to the state-of-the-art coded elastic computing scheme.
\end{abstract}


%

\section{Introduction}
Cloud computing services such as Amazon EC2 Spot and Microsoft Azure Batch offer non-dedicated computing nodes at a much
lower cost than dedicated nodes.  
The caveat of non-dedicated nodes is that they can be preempted at a short notice. Similarly, additional node may be made available. 
These two types of events are referred to as elastic events. 
Elasticity presents novel challenges in allocating tasks within the available nodes. 
The challenges due to elastic events share a thematic connection to the stragglers' problem, in which response times of nodes are unpredictable. However, there are two major properties that differentiate elastic nodes from stragglers. First, in an elastic setup new nodes can join the computation, while in the case of stragglers there is no newly available nodes that can benefit the system. Secondly, elasticity occurs with short notice which gives the master the opportunity to re-allocate computation among the available nodes. However, nodes can become stragglers without any notice.  


Recently, in~\cite{yang2019coded} a coded elastic computing (CEC) scheme was proposed that makes use of coded computing to deal with elastic events. Coded computing, which employs ideas from coding theory in parallel systems, was initially developed to address the issue of stragglers in distributed machine learning and data analytics~\cite{lee:18, yu:17, dutta2019optimal}. Coded computing introduces redundancy in computation, so that the distributed system needs only wait for a subset of nodes before recovering the output. While coded computing goes beyond simple and traditional replication that can result extremely high redundancy, the computation overhead of many of the initial coded computing designs was not optimal. To reduce computation overhead, hierarchical coded computing was proposed in~\cite{kiani:18, ferdinand:isit18, kiani:cwit19, kiani:icml19,  kiani:20} that enables both fast and slow nodes to contribute to the output recovery. In hierarchical coding, the computation completed by stragglers is exploited rather than being ignored. Building upon ~\cite{yang2019coded}, in~\cite{dau2019optimizing} a new performance criterion was introduced called \emph{transition waste}. This quantifies the total number of subtasks that existing workers must either abandon or take on anew when an elastic event occurs. In~\cite{dau2019optimizing} a new task allocation scheme is presented that achieves zero transition waste when a worker joins or leaves. These ideas were extended to heterogeneous systems in~\cite{woolsey2020heterogeneous,woolsey2020coded}. 

In this paper, we extend the idea of hierarchical coding to elastic computing. We propose two hierarchical coded elastic computing schemes: {\em multilevel coded elastic computing} (MLCEC) and {\em bit-interleaved coded elastic computing} (BICEC). The proposed MLCEC method is built upon CEC and MLCC (multilevel coded computing) which is a special case of hierarchical coding~\cite{ferdinand:isit18, kiani:cwit19, kiani:20}. In MLCEC each worker is tasked by multiple encoded subtasks and available workers select only a subset of their subtasks to work on. The task selection of MLCEC is motivated by the sequential behavior of workers. Workers process their subtasks one-by-one, transmitting each to the master as soon as it is completed. Therefore more workers are expected to finish their first selected subtasks rather than their second, and so forth. With respect to this sequential behavior, in MLCEC fewer workers select their first subtasks to work on and more workers select their last subtasks. This hierarchical selection of subtasks can make the completion time of different subtasks closer to each other and reduces the completion delay in MLCEC compared to CEC. The proposed BICEC method is built upon BICC (bit-interleaved coded computing) which is another special case of hierarchical coding~\cite{kiani:18, kiani:20}. {Similar to BICC,} in BICEC we jointly encode all subtasks using a single code and assign each worker multiple encoded subtasks. In opposite to CEC and MLCEC, where workers select a subset of their subtasks to work on, in BICEC available workers can work on all their subtasks, starting from the first one, completing subtasks sequentially until enough number of subtasks is completed. This enables a more continuous completion process compared to CEC and MLCEC. When an elastic event occurs, there is no need for task allocation, or in other words, BICEC achieves  zero transition waste. {We can also view BICEC as a BICC scheme with a higher rate of redundancy.  More redundancy in BICEC leads to robustness against more number of preempted workers.} The rest of the paper is organized as follows. In Sec.~\ref{hcec} we first provide a motivating example to illustrate the intuition behind our schemes. We then detail our proposed methods. {In Sec.~\ref{sim} we show that our method gains a $45\%$ improvement in average finishing time.} 

\section{Proposed Method}\label{hcec}
\begin{figure*}[ht]
 \centering
 \includegraphics[width=0.95\textwidth]{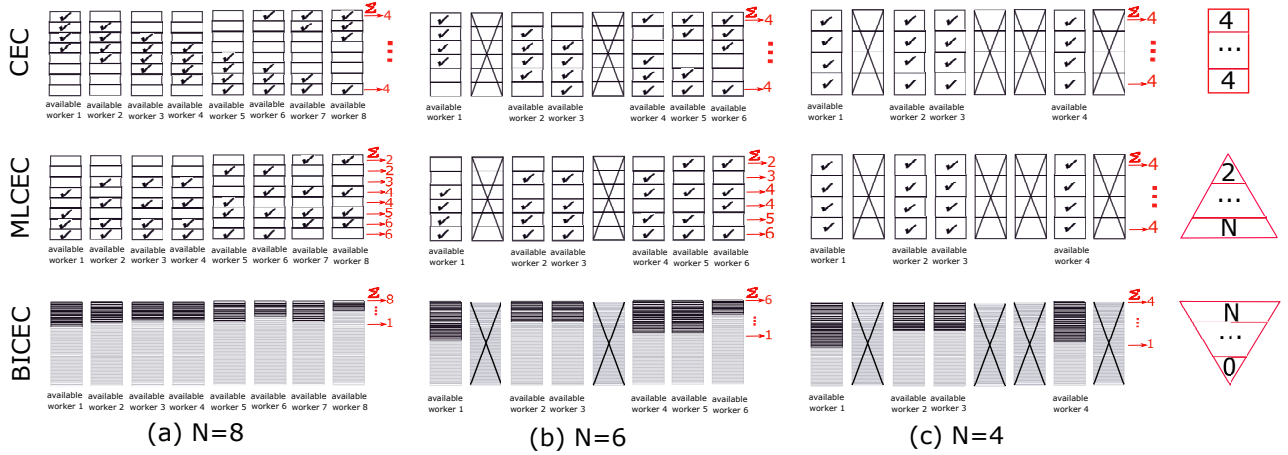}
 \vspace{-2ex}
 \caption{We plot TASs for CEC in the 1st row, for MLCEC in the 2nd row, and for BICEC in the 3rd row. Tasks are distributed among at most (a) $N=8$ workers. Workers can be preempted to be reduced to (b) $N=6$, or (c) $N=4$. The subtasks that are selected by TAS are marked as checks boxes and preempted workers are marks as cross marks. In CEC and MLCEC $K_{ece}=K_{\rm mlcec}=2$ and each worker subdivides its computation into $N$ subtasks. In BICEC  $K_{\rm bicec}=600$ and $S_{\rm bicec}=300$.} 
 \label{cec} \vspace*{-2.5ex}
\end{figure*}
Consider a system that consists of a master and $N$ workers, where workers can be preempted or joint within a range $N \in
(N_{\min}, N_{\max})$. This elasticity occurs with short notice. Each available worker can become a straggler without any prior notice. We aim to parallelize a computation in this system { by leveraging ideas from coded computing to reduce the uncertainty due to system elasticity and stragglers.}
{We first describe three illustrative examples.} { We then propose the generalized MLCEC and BICEC.} 
In the examples the goal is to multiply two matrices $A\in \mathbb{R}^{u \times w}$ and $B \in \mathbb{R}^{w \times v}$. Computing $AB\in \mathbf{R}^{u\times v}$ requires $uwv$ multiplication and addition operations. 
In the first example we detail CEC~\cite{yang2019coded}. 
In the second and third examples we present MLCEC and BICEC. 
These examples are visualized in Fig.~\ref{cec} for $N \in \{4,6,8\}$.

\textbf{Example 1 (CEC)~\cite{yang2019coded}:} We first divide the computation of the product $AB$ into $K_{\rm cec}=2$ matrix products $A_1B$ and $A_2B$. We accomplish this by partitioning $A$ horizontally into two equal-sized submatrices $A_1,A_2\in \mathbb{R}^{u/2\times w}$. We then encode these submatrices using polynomial codes~\cite{yu:17} to generate $8$ encoded submatrices $\hat{A}_n = A_1 + nA_2$, $n\in [8]$. We use the notation $[m]$ to denote the index set $\{1,\dots,m\}$. The multiplication of $\hat{A}_n$ and $B$ is assigned to worker $n\in [N]$. Note that the encoded tasks $\hat{A}_nB=A_1B + nA_2B$ can be viewed as a polynomial of degree 1. The completion of any 2 of the encoded products is enough to recover the coefficients of the polynomial, i.e., $A_1B$ and $A_2B$ products. If $N$ workers are available, then available workers subdivide their task into $N$ equal-sized subtasks. We accomplish this by horizontally dividing encoded matrices $\hat{A}_n$ into $N$ equal-sized submatrices $\hat{A}_{n,m}$, $m\in [N]$, and assigns the subtasks $\hat{A}_{n,1}B,\ldots \hat{A}_{n,N}B$ to worker $n\in[N]$. To recover $A_1B$ and $A_2B$ products, it is required to complete at least 2 computations from each of the following $N$ sets $\{\hat{A}_{n,1}B\}_{n\in [N]},\ldots,\{\hat{A}_{n,N}B\}_{n\in [N]}$. In CEC, each worker selects only $S_{\rm cec}=4$ of its $N$ subtasks to work on. The value of $4$ is selected intentionally to be larger than $K_{\rm cec}=2$ to make the system robust to stragglers. As is shown in the first row of Fig.~\ref{cec}, in CEC available workers select their to-do list of subtasks in a cyclic fashion. For example, if $N=8$ as it is in the first row of Fig.~\ref{cec}a, worker $n$ subdivides its $\hat{A}_n B$ computation into $8$ equal-sized sub-computations $\hat{A}_{n,1}B,\ldots, \hat{A}_{n,8}B$. Worker $n$ works on only $4$ subtasks, denoted by $A_{n,m}B$, where $m\equiv (n+i-1) \mod 8$ and $i \in [4]$. The cyclic allocation of CEC allows only $4$ workers to contribute to the completion of each set $\{\hat{A}_{n,m}\}_{n\in [N]}$, $m\in[8]$. In Figures~\ref{cec}b and 1c, it is shown how to continue the computation when workers are gradually preempted reducing the number of workers from $N=8$ to $N=6$ and to $N=4$. CEC uses a fixed rate code (here $2$ out of $4$) for each set recovery. Since workers complete their subtasks sequentially, the selected subtasks in the set $\{\hat{A}_{n,1}\}_{n\in [N]}$ are started to be completed sooner than the selected subtasks in the set $\{\hat{A}_{n,N}\}_{n\in [N]}$. Therefore, the completion of different sets can finish at different times. This may be wasteful of time.

\textbf{Example 2 (MLCEC): } 
The main difference of MLCEC from CEC is that each worker selects its $S_{\rm mlcec}=4$ subtasks in a way that more workers can contribute to the completion of sets $\{\hat{A}_{n,m}\}_{n\in [N]}$ with a larger $m$ when compared to the sets with a smaller $m$. While CEC selects a fixed number of workers (i.e., $4$) to contribute to the recovery of each set, MLCEC assigns different number of workers to complete each set.
For example, Fig.~\ref{cec}a shows that MLCEC assigns $d_1=d_2=2$ workers to the 1st and 2nd sets. 
However, it assigns $d_3=3$ workers to the 3rd sets. 
For 4, 5, 6, 7 and 8th sets, we use $d_4=4, d_5=4, d_6=5, d_7=6$ and $d_8=6$ workers. 
Note that while different numbers of workers contribute to each set, each worker has still $4$ subtasks to do because $\sum_{n\in [N]} d_n = 4N$. 
Similar to CEC, to complete each set, we require at least $2$ completed subtasks for that set. 
This makes the decoding cost of MLCEC similar to that of CEC. 
As the second row of Fig.~\ref{cec} shows, we set $d_1\leq d_2\leq\ldots d_N$. This setting is expected to improve the computation time since more workers can contribute to the recovery of the sets $\{\hat{A}_{n,m}\}_{n\in [N]}$ with a larger $m$, which are started later than the sets with a smaller $m$. 

\textbf{Example 3 (BICEC): } 
In the 3rd row of Fig.~\ref{cec} we illustrate how BICEC can be applied to the same example. In contrast to CEC and MLCEC, where the main computational job is divided into $2$ tasks, in BICEC we divide the job into $K_{\rm bicec}=600$ tiny computations. 
These computations are then encoded to generate $1200$ encoded subtasks. 
In Fig.~\ref{cec}a, these $1200$ encoded subtasks are distributed among $N=8$ workers so that each worker is tasked with $S_{\rm bicec}=300$ subtasks. 
The completion of any $600$ out of $1200$ computations leads to the output recovery. 
The workers start completing their tasks sequentially from the first to the $300$th. In Fig.~\ref{cec} one can see that when $0$ to $4$ workers are gradually preempted, on average only the first $y$ percentage, $y \in \{25, 33, 50\}$, of subtasks by each worker is required to be completed. 
If stragglers/failures exist, then the fast workers must complete a larger proportion of the overall computation. 
We can view the completion process of BICEC as a hierarchical process. For example, Fig.~\ref{cec}a shows that $8$ workers complete their first subtasks, while only $1$ worker completes its $90$th subtask.

\textbf{Generalizing MLCEC and BICEC: } 
We now generalize MLCEC and BICEC.
Consider a computing job $g(x)$ where $x$ is the input data and $g(.)$ is a function. For any positive integer $k$, let's assume that the job $g(x)$ can be decomposed into $k$ computations, i.e., $g(x) = f_k(g_1(x),\ldots,g_k(x))$, where the function $f_k(.)$ maps the $k$ computations $g_1(x),\ldots,g_k(x)$ to $g(x)$. We also assume that the $g_1(x),\ldots, g_k(x)$ computations are linear, i.e., for any $i,j \in [k]$, $ag_i(x)+bg_j(x) = (ag_1+bg_2)(x)$. One example is a matrix multiplication $g(x)=Ax$. The input $x$ is a $u\times w$ matrix and $A$ is a $w \times v$ matrix. The job $Ax$ can then be partitioned into $k$ computations $g_i(x)=A_ix$, $i \in [k]$, by horizontally dividing $A$ into $k$ equal-sized submatrices $A_i$. The function $f_k(.)$ simply concatenates the $A_ix$ results, for all $i \in [k]$. Note that if the total number of computations is not divisible by $k$, we can use zero-padding. 

\noindent {\textit{\textbf{Multilevel coded elastic computing:}}} 
The master first divides the job $g(x)$ into $K_{\rm mlcec}$ computations $g_{i}(x)$, where $i \in [K_{\rm mlcec}]$. I.e., $g(x)=f_{K_{\rm mlcec}}(g_1(x),\ldots,g_{K_{\rm mlcec}}(x))$. These computations are then encoded using an $(K_{\rm mlcec},N_{\max})$ MDS code~\cite{lee:18} to generate $N_{\max}$ encoded tasks $\hat{g}_n(x)$, $n \in [N_{\max}]$. Note that the completion of any $K_{\rm mlcec}$ encoded tasks can lead to $g(x)$ recovery. The master then assigns the encoded task $\hat{g}_n(x)$ to the $n$th worker, for all $n \in [N_{\max}]$. If $N$ workers are available, each worker subdivides its tasks into $N$ equal-sized subtasks. Let $\hat{g}_{n}^{1}(x),\ldots,\hat{g}_{n}^{N}(x)$ be the $N$ subtasks of the $n$th available worker, $n \in [N]$. Let's assume that these $N^2$ encoded subtasks $\hat{g}_{n}^{m}(x)$, $n,m \in [N]$, can be reconstructed differently. The master can first divide the job $g(x)$ into $NK_{\rm mlcec}$ computations $g_{i}^{m}(x)$, where $i \in [K_{\rm mlcec}]$ and $m \in [N]$. I.e., $g(x)=f_{NK_{\rm mlcec}}(g_{1}^{1}(x),\ldots,g_{K_{\rm mlcec}}^{N}(x))$. The master then groups these computations into $N$ sets $\{g_{i}^{m}\}_{i\in [K_{\rm mlcec}]}$, where $m\in [N]$. The computations of each set is then encoded using an $(K_{\rm mlcec},N)$ MDS code~\cite{lee:18} to generate $N$ encoded subtasks. For the set $\{\hat{g}_{n}^m(x)\}_{n \in [N]}$, $m \in [N]$, the completion of any $K_{\rm mlcec}$ encoded subtasks can recover all $g_{i}^{m}(x)$, $i \in [K_{\rm mlcec}]$. To reduce redundant computation, in MLCEC each worker selects only $S_{\rm mlcec}$ of its subtasks to work on. Also, assume that $d_m$ workers select their $m$th encoded subtask, $m \in [N]$. Note that the $m$th encoded substask of each worker belongs to the set $\{\hat{g}_{n}^m(x)\}_{n \in [N]}$. By double counting, we have $\sum_{m\in [N]} d_m = S_{\rm mlcec}N$.
 Since each worker processes its selected subtasks sequentially, fewer workers are expected to finish their last selected subtasks compared to their first ones. To make the chance of completion of all sets $\{\hat{g}_{n}^m(x)\}_{n \in [N]}$, $m \in [N]$, closer to each other, we set $d_1 \leq \ldots \leq d_N$. While due to space constraint, we must leave discussion of how to optimize the set $\{d_m\}_{m\in [N]}$ to future work, in Alg.~\ref{alg} we describe one method to allocate selected subtasks given $\{d_m\}_{m\in [N]}$.  
\begin{algorithm}  \vspace*{-3.5ex}
 \KwData{$N, \{d_1,\ldots,d_N\}$}
 All workers are initiated with 0 subtasks\;
 \For{$l$ = $N$ to $1$}{
  $n=$ index of the 1st worker who has the minimum number of subtasks in sets $l+1$ to $N$\;
  \For{$i$ = $n$ to $n+d_l$}{
    worker $i\mod N$ selects its $l$th subtask\; 
  }
 }
 \caption{Task allocation in MLCEC}\label{alg} 
\end{algorithm}
\vspace*{-3ex}

\noindent {\textit{\textbf{Bit-interleaved coded elastic computing:}}} 
The master divides the job $g(x)$ into $K_{\rm bicec}$ computations ${g}_{i}(x)$, i.e., $g(x) = f_{K_{\rm bicec}}({g}_1(x), \ldots, {g}_{K_{\rm bicec}}(x))$. These ${g}_{i}(x)$ are then encoded using an $(K_{\rm bicec},S_{\rm bicec}N_{\max})$ MDS code~\cite{lee:18} to generate $S_{\rm bicec}N_{\max}$ encoded subtasks $\hat{g}_{i}(x)$, $i \in [S_{\rm bicec}N_{\max}]$. The master then assigns $S_{\rm bicec}$ encoded subtasks to each worker. The completion of any $K_{\rm bicec}$ encoded subtasks can recover all ${g}_i(x)$, $i \in [K_{\rm bicec}]$. Workers process their subtasks sequentially until enough subtasks are completed. 
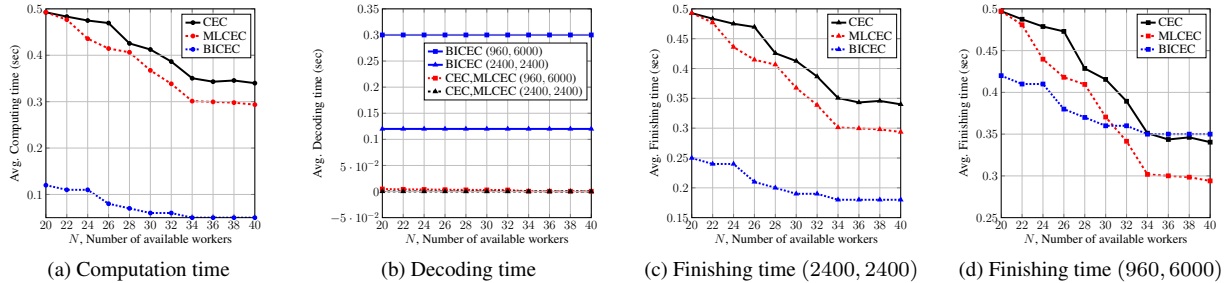
\begin{figure*}[ht]
\centering 
		\subfloat[Computation time]{\tikzset{every mark/.append style={scale=0.4}}
	\input{./image/fig_comp_vs_N.tex}
 \label{FIG:comp}}
	\quad%
	\subfloat[Decoding time]{\tikzset{every mark/.append style={scale=0.4}}
	\input{./image/fig_dec_vs_N.tex}
 \label{FIG:dec}}
	\quad%
		\subfloat[Finishing time $(2400,2400)$]{\tikzset{every mark/.append style={scale=0.4}}
	\input{./image/fig_fin_vs_N.tex}
 \label{FIG:fin}}
	\quad%
		\subfloat[Finishing time $(960,6000)$]{\tikzset{every mark/.append style={scale=0.4}}
	   	\input{./image/fig_fin1_vs_N.tex}
 \label{FIG:fin1}}
	\quad
	\setlength{\belowcaptionskip}{-12pt}
	 \vspace{-1.5ex}
\caption{For BICEC, MLCEC, and CEC we plot (a) Average computation time vs. $N$, where $uwv=2400^3$. (b) Average decoding time vs. $N$ in two cases when $(u,w,v)=(2400,960,6000)$ and when $(u,w,v)=(2400,2400,2400)$. (c),(d) Average finishing time vs. $N$ where in (c) $(u,w,v)=(2400,2400,2400)$ and in (d) $(u,w,v)=(2400,960,6000)$. In all subfigures, we set $K_{\rm cec}=K_{\rm mlcec}=10, K_{\rm bicec}=800, S_{\rm cec}=S_{\rm mlcec}=20,S_{\rm bicec}=80,$ and $N_{\max}=40$.}
\label{vs_N} \vspace*{-2.5ex}
\end{figure*}

\section{Simulation results}\label{sim}
\vspace{-0.5em}
We now evaluate the performance of MLCEC and BICEC and compare our results with CEC~\cite{yang2019coded}. We conduct our experiments using {\em Python} on a machine with an {\em Intel(R) Core(TM) i7-7700HQ CPU @ 2.80GHz}. We simulate the matrix multiplication $AB$ on $N$ working nodes, where $N \in \{20,22,\ldots,40\}$. 
We run the computation assigned to workers sequentially (worker-by-worker). 
The parallelism among workers' computations is simulated by recording workers' computation times and measuring the final computation times for CEC, MLCEC, and BICEC. To simulate straggling behavior, each available worker becomes straggler with probability $0.5$. 
After computation completes, we implement a decoding process and log the decoding time. 
We repeat our experiments twice, once for multiplying two square random matrices $A\in \mathbb{R}^{2400\times 2400}$ and $B \in \mathbb{R}^{2400\times 2400}$, and another time for multiplying a tall matrix $A \in \mathbb{R}^{2400\times 960}$ and a fat matrix $B \in \mathbb{R}^{960 \times 6000}$. For each case, we run our simulation $20$ times. In  Fig.~\ref{vs_N} we plot the average decoding, computation, and finishing (computation plus decoding) times versus $N$. We do not report encoding time as it is negligible when compared to computation times. This is because encoding requires many fewer multiplication and addition operations compared to the at least $uwv=2400^3$ operations done by workers. We also neglect the time it takes to communicate data between workers and the master.



For CEC and MLCEC, we set $K_{\rm cec}=K_{\rm mlcec}=10$. Since $N_{\max}=40$ 
, in our experiments we generate $40$ random matrices $\hat{A}_n \in \mathbb{R}^{\frac{u}{10} \times w}$, $n\in [40]$, each of which corresponds to an encoded data input for one worker. We also generate a random matrix $B$. We then assign the $\hat{A}_nB$ multiplication to worker $n \in [N]$. This multiplication requires $\frac{2400^3}{10}$ multiplication and addition operations. Each worker then subdivides its task into $N$ equal-sized subtasks and works on $20$ subtasks according to the task allocation schemes in CEC and MLCEC. For each value of $N$, we set $\{d_n\}_{n\in [N]}$ such that $d_1 \leq \ldots \leq d_N$ and $\sum_{n \in [N]} d_n = 20 N$. For BICEC, we set $K_{\rm bicec}=800$ and $S_{\rm bicec}=80$. Therefore, each worker is tasked by at most $\frac{2400^3}{10}$ computations, similar to CEC and MLCEC. In Fig.~\ref{FIG:comp}, we plot the average computation time vs. $N$ for CEC, MLCEC, and BICEC. The computation time for both matrix dimensions $(u,w,v)=(2400,2400,2400)$ and $(u,w,v)=(2400,960,6000)$ are the same because $uwv=2400^3$ for both cases. In Fig.~\ref{FIG:comp}, MLCEC achieves a lower average computation time compared to CEC as we expected. BICEC has the lowest average computation time and achieves $85\%$ improvement compared to CEC for $N=40$. 

For decoding, we solve a system of linear equations that involves a Vandermonde matrix of size $10 \times 10$ in CEC and MLCEC and of size $800\times 800$ in BICEC. After we take the inverse of the Vandermonde matrix, in CEC and MLCEC $\frac{10uv}{N}$ multiplication and addition operations are completed. In BICEC $800uv$ operations are completed. Therefore, as is shown in Fig.~\ref{FIG:dec}, BICEC has the worst decoding time and CEC and MLCEC both have the same (negligible) decoding time. If we change from $(u,w,v)=(2400,2400,2400)$ to $(u,w,v)=(2400,960,6000)$ the decoding time increases since the decoding process only depends on the dimensions of $AB$, which is $u\times v$, and $v$ in the first setting is smaller than in the latter. In Fig.~\ref{FIG:fin} and~\ref{FIG:fin1} we plot the average computation plus decoding time (denoted ``finishing time''). In Fig.~\ref{FIG:fin} BICEC is the best choice and in Fig.~\ref{FIG:fin1} MLCEC has the lowest finishing time for $N\in \{32,\ldots,40\}$. While in both sub-figures computation time is fixed, in Fig.~\ref{FIG:fin1} the large decoding time in BICEC is added to the computation time and increases the finishing time of BICEC. For $N=40$, the use of BICEC yields a $45\%$ reduction in average finishing time when compared to CEC in Fig.~\ref{FIG:fin} and the use of MLCEC yields a $15\%$ reduction in Fig.~\ref{FIG:fin1}.

BICEC has two advantages over MLCEC at the cost of larger decoding time. Due to a more continuous completion process of BICEC, its computation time is a lower bound for MLCEC. Besides this, whenever an elastic event occurs, in MLCEC available workers may need to re-allocate their subtasks, while in BICEC workers keep working on their pre-allocated subtasks, resulting in zero transition waste. 
{As next steps, we would like to implement our schemes on real-world elastic computing frameworks such as Amazon EC2 Spot or Microsoft Azure Batch, and to extend our schemes to a larger class of algorithms rather than linear computations.}



\clearpage

\bibliographystyle{IEEEbib}
\bibliography{IEEEabrv,references}

\end{document}

%% file: image/fig_comp_vs_N.tex
	\begin{tikzpicture}[scale=0.33]
	\begin{axis}[
	height=10cm,
	width=10cm,
	grid=major,
	xlabel={\Large $N$, Number of available workers},
	ylabel={\Large Avg. Computing time (sec)},
legend style={at={(0.65,0.87)},anchor=west,nodes=right},	y tick label style={font=\Large},
    x tick label style={font=\Large},
	axis on top,xmin=20, xmax=40, ymin=0.05, ymax=0.5]

		\addlegendentry{\Large CEC}
		\addplot [line width=0.8mm, color=black, solid, every mark/.append style={solid, fill=black},mark=otimes*, mark size=4pt] coordinates {
		(20,0.4925)
		(22,0.48329)
		(24,0.47485)
		(26,0.4694)
		(28,0.4255) 
		(30,0.41241)	
		(32,0.3864)
		(34,0.3505)
		(36,0.34302)
		(38,0.3456) 
		(40,0.3399)	   
	};

	\addlegendentry{\Large MLCEC}
		\addplot [line width=0.8mm, color=red, dashed, every mark/.append style={solid, fill=red},mark=otimes*, mark size=4pt] coordinates {
		(20,0.4925)
		(22,0.4767)
		(24,0.43576)
		(26,0.41449)
		(28,0.4064) 
		(30,0.3674)	
		(32,0.33851)
		(34,0.30119)
		(36,0.2995)
		(38,0.2979) 
		(40,0.29353)	 	
	};

	\addlegendentry{\Large BICEC}
	\addplot [line width=0.8mm, color=blue, dotted, every mark/.append style={solid, fill=blue},mark=otimes*, mark size=4pt] coordinates {
		(20,0.12)
		(22,0.11)
		(24,0.11)
		(26,0.08)
		(28,0.07) 
		(30,0.06)	
		(32,0.06)
		(34,0.05)
		(36,0.05)
		(38,0.05) 
		(40,0.05)	
 	};

	\end{axis}
	\end{tikzpicture} 

%% file: image/fig_dec_vs_N.tex
	\begin{tikzpicture}[scale=0.33]
	\begin{axis}[
	height=10cm,
	width=10cm,
	grid=major,
	xlabel={\Large $N$, Number of available workers},
	ylabel={\Large Avg. Decoding time (sec)},
legend style={at={(0.2,0.7)},anchor=west,nodes=right},	y tick label style={font=\Large},
    x tick label style={font=\Large},
	axis on top,xmin=20, xmax=40, ymin=-0.05, ymax=0.35]

		\addlegendentry{\Large BICEC ($960, 6000$) }
		\addplot [line width=0.8mm, color=blue, solid, every mark/.append style={solid, fill=blue},mark=square*, mark size=4pt] coordinates {
		(20,0.3)
		(22,0.3)
		(24,0.3)
		(26,0.3)
		(28,0.3) 
		(30,0.3)	
		(32,0.3)
		(34,0.3)
		(36,0.3)
		(38,0.3) 
		(40,0.3)	   
	};

	\addlegendentry{\Large BICEC ($2400,2400$) }
		\addplot [line width=0.8mm, color=blue, solid, every mark/.append style={solid, fill=blue},mark=triangle*, mark size=4pt] coordinates {
		(20,0.12)
		(22,0.12)
		(24,0.12)
		(26,0.12)
		(28,0.12) 
		(30,0.12)	
		(32,0.12)
		(34,0.12)
		(36,0.12)
		(38,0.12) 
		(40,0.12)	 	
	};

	\addlegendentry{\Large CEC,MLCEC ($960, 6000$)}
	\addplot [line width=0.8mm, color=red, dotted, every mark/.append style={solid, fill=red},mark=square*, mark size=4pt] coordinates {
		(20,0.0048)
		(22,0.00424)
		(24,0.00399)
		(26,0.00359)
		(28,0.00316) 
		(30,0.0031)	
		(32,0.00295)
		(34,0.00065)
		(36,0.00058)
		(38,0.00056) 
		(40,0.00049)	
 	}; 	
 	
 		\addlegendentry{\Large CEC,MLCEC ($2400, 2400$) }
	\addplot [line width=0.8mm, color=black, dashed, every mark/.append style={solid, fill=black},mark=triangle*, mark size=4pt] coordinates {
		(20,0.00033)
		(22,0.00025)
		(24,0.0002)
		(26,0.00018)
		(28,0.00015) 
		(30,0.00012)	
		(32,0.00012)
		(34,0.00012)
		(36,0.0001)
		(38,0.00009) 
		(40,0.00009)	
 	};

	\end{axis}
	\end{tikzpicture} 

%% file: image/fig_fin_vs_N.tex
	\begin{tikzpicture}[scale=0.33]
	\begin{axis}[
	height=10cm,
	width=10cm,
	grid=major,
	xlabel={\Large $N$, Number of available workers},
	ylabel={\Large Avg. Finishing time (sec)},
legend style={at={(0.65,0.87)},anchor=west,nodes=right},	y tick label style={font=\Large},
    x tick label style={font=\Large},
	axis on top,xmin=20, xmax=40, ymin=0.15, ymax=0.5]

		\addlegendentry{\Large CEC}
		\addplot [line width=0.8mm, color=black, solid, every mark/.append style={solid, fill=black},mark=triangle*, mark size=4pt] coordinates {
		(20,0.49283)
		(22,0.48354)
		(24,0.47505)
		(26,0.46958)
		(28,0.42565) 
		(30,0.41253)	
		(32,0.38652)
		(34,0.35062)
		(36,0.34312)
		(38,0.34569) 
		(40,0.33999)	   
	};

	\addlegendentry{\Large MLCEC}
		\addplot [line width=0.8mm, color=red, dashed, every mark/.append style={solid, fill=red},mark=triangle*, mark size=4pt] coordinates {
		(20,0.49283)
		(22,0.47695)
		(24,0.43596)
		(26,0.41467)
		(28,0.40655) 
		(30,0.36752)	
		(32,0.33863)
		(34,0.30131)
		(36,0.2996)
		(38,0.29799) 
		(40,0.29362)	 	
	};

	\addlegendentry{\Large BICEC}
	\addplot [line width=0.8mm, color=blue, dotted, every mark/.append style={solid, fill=blue},mark=triangle*, mark size=4pt] coordinates {
		(20,0.25)
		(22,0.24)
		(24,0.24)
		(26,0.21)
		(28,0.2) 
		(30,0.19)	
		(32,0.19)
		(34,0.18)
		(36,0.18)
		(38,0.18) 
		(40,0.18)	
 	};

	\end{axis}
	\end{tikzpicture} 

%% file: image/fig_fin1_vs_N.tex
	\begin{tikzpicture}[scale=0.33]
	\begin{axis}[
	height=10cm,
	width=10cm,
	grid=major,
	xlabel={\Large $N$, Number of available workers},
	ylabel={\Large Avg. Finishing time (sec)},
legend style={at={(0.65,0.87)},anchor=west,nodes=right},	y tick label style={font=\Large},
    x tick label style={font=\Large},
	axis on top,xmin=20, xmax=40, ymin=0.25, ymax=0.5]

		\addlegendentry{\Large CEC}
		\addplot [line width=0.8mm, color=black, solid, every mark/.append style={solid, fill=black},mark=square*, mark size=4pt] coordinates {
		(20,0.49718)
		(22,0.48753)
		(24,0.47884)
		(26,0.47299)
		(28,0.42866) 
		(30,0.41551)	
		(32,0.38935)
		(34,0.35115)
		(36,0.3436)
		(38,0.34616) 
		(40,0.34039)	   
	};

	\addlegendentry{\Large MLCEC}
		\addplot [line width=0.8mm, color=red, dashed, every mark/.append style={solid, fill=red},mark=square*, mark size=4pt] coordinates {
		(20,0.49718)
		(22,0.48094)
		(24,0.43975)
		(26,0.41808)
		(28,0.40956) 
		(30,0.3705)	
		(32,0.34146)
		(34,0.30184)
		(36,0.30008)
		(38,0.29846) 
		(40,0.29402)	 	
	};

	\addlegendentry{\Large BICEC}
	\addplot [line width=0.8mm, color=blue, dotted, every mark/.append style={solid, fill=blue},mark=square*, mark size=4pt] coordinates {
		(20,0.42)
		(22,0.41)
		(24,0.41)
		(26,0.38)
		(28,0.37) 
		(30,0.36)	
		(32,0.36)
		(34,0.35)
		(36,0.35)
		(38,0.35) 
		(40,0.35)	
 	};

	\end{axis}
	\end{tikzpicture} 

%% file: main.bbl
\begin{thebibliography}{10}

\bibitem{yang2019coded}
Yaoqing Yang, Matteo Interlandi, Pulkit Grover, Soummya Kar, Saeed Amizadeh,
  and Markus Weimer,
\newblock ``Coded elastic computing,''
\newblock in {\em IEEE Int. Symp. Inf. Theory (ISIT)}, July 2019.

\bibitem{lee:18}
Kangwook Lee, Maximilian Lam, Ramtin Pedarsani, Dimitris Papailiopoulos, and
  Kannan Ramchandran,
\newblock ``Speeding up distributed machine learning using codes,''
\newblock {\em IEEE Trans. on Inf. Theory}, vol. 64, pp. 1514--1529, Mar. 2018.

\bibitem{yu:17}
Qian Yu, Mohammad Maddah-Ali, and Salman Avestimehr,
\newblock ``Polynomial codes: An optimal design for high-dimensional coded
  matrix multiplication,''
\newblock in {\em Int. Conf. Neural Inf. Proc. Sys. (NIPS)}, 2017.

\bibitem{dutta2019optimal}
Sanghamitra Dutta, Mohammad Fahim, Farzin Haddadpour, Haewon Jeong, Viveck
  Cadambe, and Pulkit Grover,
\newblock ``On the optimal recovery threshold of coded matrix multiplication,''
\newblock {\em IEEE Trans. on Inf. Theory}, vol. 66, no. 1, pp. 278--301, 2019.

\bibitem{kiani:18}
Shahrzad Kiani, Nuwan Ferdinand, and Stark~C Draper,
\newblock ``Exploitation of stragglers in coded computation,''
\newblock in {\em IEEE Int. Symp. Inf. Theory (ISIT)}, 2018.

\bibitem{ferdinand:isit18}
Nuwan Ferdinand and Stark~C Draper,
\newblock ``Hierarchical coded computation,''
\newblock in {\em IEEE Int. Symp. Inf. Theory (ISIT)}, 2018.

\bibitem{kiani:cwit19}
Shahrzad Kiani, Nuwan Ferdinand, and Stark~C Draper,
\newblock ``Hierarchical coded matrix multiplication,''
\newblock in {\em Canadian Workshop on Inf. Theory}, 2019.

\bibitem{kiani:icml19}
Shahrzad Kiani, Nuwan Ferdinand, and Stark~C Draper,
\newblock ``Cuboid partitioning for hierarchical coded matrix multiplication,''
\newblock {\em IEEE Int. Conf. on Machine Learning (Workshop on Coded Machine
  Learning), preprint arXiv: 1907.08819}, 2019.

\bibitem{kiani:20}
Shahrzad Kiani, Nuwan Ferdinand, and Stark~C Draper,
\newblock ``Hierarchical coded matrix multiplication,''
\newblock {\em IEEE Trans. Inf. Theory}, vol. 67, no. 2, pp. 726--754, 2021.

\bibitem{dau2019optimizing}
Hoang Dau, Ryan Gabrys, Yu-Chih Huang, Chen Feng, Quang-Hung Luu, Eidah
  Alzahrani, and Zahir Tari,
\newblock ``Optimizing the transition waste in coded elastic computing,''
\newblock in {\em IEEE Int. Symp. Inf. Theory (ISIT)}, 2020, pp. 174--178.

\bibitem{woolsey2020heterogeneous}
N.~{Woolsey}, R.~R. {Chen}, and M.~{Ji},
\newblock ``Heterogeneous computation assignments in coded elastic computing,''
\newblock in {\em IEEE Int. Symp. Inf. Theory (ISIT)}, 2020, pp. 168--173.

\bibitem{woolsey2020coded}
Nicholas Woolsey, Rong-Rong Chen, and Mingyue Ji,
\newblock ``Coded elastic computing on machines with heterogeneous storage and
  computation speed,''
\newblock {\em arXiv preprint arXiv:2008.05141}, 2020.

\end{thebibliography}
